\documentclass{amsproc}
\usepackage{amssymb}

\usepackage{amsmath}
\usepackage{graphicx}

\theoremstyle{definition}

\theoremstyle{remark}

\numberwithin{equation}{section}

\input{tcilatex}

\begin{document}
\title{Nambu Dynamics, Deformation Quantization, and Superintegrability}
\author{Thomas L Curtright}
\address{Department of Physics, University of Miami, Coral Gables, Florida
33124-8046}
\email{curtright@physics.miami.edu}
\author{Cosmas K Zachos}
\address{High Energy Physics Division, Argonne National Laboratory, Argonne,
Illinois 60439-4815}
\email{zachos@hep.anl.gov}
\thanks{The first author was supported in part by NSF Award 0073390.}
\thanks{The second author was supported in part by the US Department of
Energy, Division of High Energy Physics, Contract W-31-109-ENG-38.}
\date{January 1, 1994 and, in revised form, June 22, 1994.}
\dedicatory{This paper is dedicated to Milton Hendrix (1930-2000).}

\begin{abstract}
Phase space is a framework ideally suited for quantizing superintegrable
systems through the use of deformation methods, as illustrated here by
applications to de Sitter and chiral particles. \ Within this framework,
Nambu brackets elegantly incorporate the additional quantum invariants of
such models. \ New results are presented for the non-Abelian quantization of
these brackets.
\end{abstract}

\maketitle

\section{Introduction\protect\footnote{%
Presented by the first author at the Workshop on Superintegrability in
Classical and Quantum Systems, Centre de recherches math\'{e}matiques,
Universit\'{e} de Montr\'{e}al, 16-21 Sept 2002.}}

For systems with velocity-dependent potentials, when quantization of the
classical system presents operator ordering ambiguities involving $x$ and $p$%
, the general consensus has long been \cite{winter,velo,lakshmanan, higgs,
leemon} to select those orderings in the quantum Hamiltonian which maximally
preserve the symmetries present in the corresponding classical Hamiltonian.
\ However, even for simple systems such constructions may become involved
and quite technical.

Recently it was emphasized \cite{CurtrightZachos,ZachosCurtright} that in
contrast to conventional operator quantization, the problem of selecting the
quantum Hamiltonian which maximally preserves integrability is addressed
very directly in Moyal's phase-space quantization formulation \cite%
{moyal,cfz,czreview}. \ Basically, the reason for this is that the kernels
involved (``kernel functions'' or ``Weyl transforms of operators'') are
c-number functions and have an interpretation analogous to that of the
classical phase-space theory, although in general they involve $\hbar $%
-corrections (``deformations''). \ This stand-alone formulation of quantum
mechanics is based on the Wigner Function (WF), a quasi-probability
distribution function in phase-space which comprises the kernel function of
the density matrix. \ Observables and transition amplitudes are phase-space
integrals of kernel functions weighted by the WF, in analogy to statistical
mechanics. \ 

There is no free lunch, however. \ Kernel functions compose through the $%
\star$-product, a noncommutative, associative, pseudodifferential operation,
which encodes the entire quantum mechanical action and whose
antisymmetrization (commutator) is the Moyal Bracket (MB) \cite%
{moyal,cfz,czreview}. \ Any arbitrary operator ordering can be brought to
Weyl-ordering format, by use of Heisenberg commutations, and through Weyl's
transform corresponds invertibly to a specific $\hbar$-deformation of the
classical kernel \cite{weyl,groenewold}. \ Thus, two operators of different
orderings correspond to kernel functions differing in their deformation
terms of $O(\hbar)$, and\ the problem of selecting the correct ordering
reduces to a purely $\star$-product algebraic one.

Previously, Hietarinta \cite{hietarinta} has investigated in this
phase-space quantization language the simplest integrable systems of
velocity-dependent potentials. \ In each system, he has promoted the
vanishing of the Poisson Bracket (PB) of the (one) classical invariant $I$
(conserved integral) with the Hamiltonian, $\{H,I\}=0$, to the vanishing of
its (quantum) Moyal Bracket (MB) with the Hamiltonian, $\{H_{qm},I_{qm}\}_{%
{\footnotesize MB}}=0$. \ This dictates quantum corrections, addressed
perturbatively in $\hbar$: \ he has found $O(\hbar^{2})$ corrections to the $%
I$s and $H$ ($V$), needed for quantum symmetry. \ The expressions found are
quite simple, as the systems chosen are such that the polynomial character
of the $p$s, or suitable balanced combinations of $p$s and $q$s, ensure
collapse or subleading termination of the MBs. \ The specification of the
symmetric Hamiltonian then is complete, since the quantum Hamiltonian in
terms of classical phase-space variables corresponds uniquely to the
Weyl-ordered expression for these variables in operator language. \ Berry %
\cite{berry} has also studied the WFs of integrable systems.

In this contribution we shall review the work in \cite%
{CurtrightZachos,ZachosCurtright}\ and make some extensions of it. \ We
shall discuss nonlinear $\sigma$-models to argue for the general principles
of power and convenience in isometry-preserving quantization in phase space,
particularly for superintegrable situations \cite{winter}. \ Briefly, we
find that the symmetry generator invariants are undeformed by quantization,
but the Casimir invariants of their MB algebras are deformed. \ Hence the
Hamiltonians are also deformed by terms $O(\hbar^{2})$, as they consist of
quadratic Casimir invariants, but nonetheless their energy spectra can be
read off through the usual group theoretic techniques once these are
properly adapted to phase space. \ The basic principles are illustrated for
the two-sphere, in Section 2, and then applied to larger classes of
symmetric manifolds such as $N$-spheres, in Section 3, and chiral models, in
Section 4. \ 

Finally, in Section 5, the evolution of such systems is formulated in
phase-space through the use of Nambu Brackets (NBs) \cite%
{nambu,hietnambu,takhtajan,nutku}. \ The quantization of Nambu's approach is
discussed in detail and compared to the standard Moyal deformation
quantization. \ This last section is the most original part of the paper,
and in it we present several new results involving Quantum Nambu Brackets
(QNBs). \ An Appendix indicates how the QNB results can also be expressed in
the more traditional language of operators acting on Hilbert space.

\section{Basic principles and the 2-sphere}

A Hamiltonian system with $N$ degrees of freedom is \emph{integrable} if it
has $N$ invariants in involution (globally defined and independent), and 
\emph{superintegrable} \cite{winter} if it has additional conservation laws
up to a maximum $2N-1$ invariants. \ For example, consider a particle
constrained to the surface of a unit radius two-sphere $S^{2}$, but
otherwise moving freely. \ Thus $N=2$ and $2N-1=3$. \ Three independent
invariants of this maximally superintegrable system are the angular momenta
about the center of the sphere: 
\begin{equation*}
L_{x}\;,\;\;\;L_{y}\;,\;\;\;L_{z}\;.
\end{equation*}
Actually, no two of these are in involution, but this is quickly remedied,
and moreover it isn't much of a hindrance as we shall see below in the Nambu
approach to mechanics, an approach that places all invariants on a more
equal footing.

To be more explicit, we may coordinate the upper and lower ($\pm$)
hemispheres by projecting the particle's location onto the equatorial disk, $%
\left\{ \left( x,y\right) \;|\;x^{2}+y^{2}\leq1\right\} $. \ The invariants
are then 
\begin{align*}
L_{z} & =xp_{y}-yp_{x}\;, \\
L_{y} & =\pm\sqrt{1-x^{2}-y^{2}}\;p_{x}\;, \\
L_{x} & =\mp\sqrt{1-x^{2}-y^{2}}\;p_{y}\;.
\end{align*}
The last two are the de Sitter momenta, or nonlinearly realized axial
charges, corresponding to the $x,y$ ``pions'' of this truncated $\sigma$%
-model.

The Poisson Brackets (PBs) of these expressions close into the expected $%
so(3)$. 
\begin{equation*}
\{L_{x},L_{y}\}_{{\footnotesize PB}}=L_{z}~,\qquad\{L_{y},L_{z}\}_{%
{\footnotesize PB}}=L_{x}~,\qquad\{L_{z},L_{x}\}_{{\footnotesize PB}%
}=L_{y}\;.
\end{equation*}
The usual Hamiltonian of the free particle system is the Casimir invariant. 
\begin{equation*}
H=\frac{1}{2}\left( L_{x}L_{x}+L_{y}L_{y}+L_{z}L_{z}\right) \;.
\end{equation*}
Thus, it immediately follows algebraically that PBs of $H$ with the $\mathbf{%
L}$ vanish, and time-invariance holds. 
\begin{equation*}
\frac{d}{dt}\mathbf{L}=\{\mathbf{L},H\}_{{\footnotesize PB}}=0\;.
\end{equation*}
So any one of the $L$s and this Casimir constitute a pair of invariants in
involution.

In quantizing this system using operators some simple ordering issues arise
because the system has an effective momentum-dependent potential. \ 
\begin{equation*}
H=\frac{1}{2}\left( 1-x^{2}\right) p_{x}^{2}+\frac{1}{2}\left(
1-y^{2}\right) p_{y}^{2}-xyp_{x}p_{y}\;.
\end{equation*}
But in the deformation quantization of phase-space one may just insert
Groenewold's \cite{groenewold} non-commutative but associative $\star $%
-products\footnote{%
A general discussion of star products on curved spaces may be found in \cite%
{Fedosov} and \cite{Yakimov}.}, defined as 
\begin{equation*}
\star\equiv\exp{\frac{i\hbar}{2}}(\overset{\leftarrow}{\partial}_{x}\overset{%
\rightarrow}{\partial}_{p_{x}}-\overset{\leftarrow}{\partial}_{p_{x}}\overset%
{\rightarrow}{\partial}_{x}+\overset{\leftarrow}{\partial}_{y}\overset{%
\rightarrow}{\partial}_{p_{y}}-\overset{\leftarrow}{\partial }_{p_{y}}%
\overset{\rightarrow}{\partial}_{y})\;,
\end{equation*}
at strategic points in the above classical expressions and achieve full
quantum integrability. \ That is, the classical Poisson bracket statements, 
\begin{equation*}
\{I,H\}_{{\footnotesize PB}}=0\text{ \ \ \ \ \ (for invariants)\ ,}
\end{equation*}
may be promoted to quantum Moyal bracket statements, 
\begin{equation*}
\{I_{qm},H_{qm}\}_{{\footnotesize MB}}\equiv\frac{1}{i\hbar}\left(
I_{qm}\star H_{qm}-H_{qm}\star I_{qm}\right) =0\text{ \ \ \ \ \ (for
invariants)\ ,}
\end{equation*}
with simple choices for $I_{qm}$\ and $H_{qm}$\ . \ As $\hbar\rightarrow0$,
the MB reduces to the PB. \ More importantly, the MB provides the unique
route around the Groenewold-van Hove theorem\footnote{%
Groenewold introduced the $\star$-product to evade his ``no-go'' theorem in
the same 1946 paper wherein he proved it, thereby immediately reducing the
theorem to a mere appurtenance.}. \ The $\star$ product is the unique (up to
equivalence) one-parameter associative deformation of ordinary products, and
the MB is the corresponding deformation of the PB.

Maximally symmetric phase-space quantization is achieved here by a
Hamiltonian of the form 
\begin{equation*}
H_{qm}=\frac{1}{2}\left( L_{x}\star L_{x}+L_{y}\star L_{y}+L_{z}\star
L_{z}\right) \;.
\end{equation*}
The reason is that, in this realization, there are no $O(\hbar)$ corrections
to the individual $\mathbf{L}$s. \ These particular invariants are
undeformed by quantization. 
\begin{equation*}
\mathbf{L}=\mathbf{L}_{qm}\;.
\end{equation*}
The algebra of the $\mathbf{L}$s$\mathbf{\ }$is then promoted to the
corresponding MB expression \emph{without any modification}, since all ``$%
\mathbf{L}$ with$\ \mathbf{L}$'' MBs collapse to PBs by the linearity in
momenta of the arguments. \ Consequently, given associativity for $\star$,
the corresponding quantum quadratic Casimir invariant $\mathbf{L}\cdot \star%
\mathbf{L}$ has vanishing MBs with $\mathbf{L}$, and automatically generates
a symmetry-preserving time-evolution. \ The $\star$-product in this
Hamiltonian trivially evaluates to expose a quantum correction to the
classical phase-space energy: 
\begin{equation*}
H_{qm}=H+\frac{\hbar^{2}}{8}(\frac{1}{1-x^{2}-y^{2}}-3)\;.
\end{equation*}
Thus we have 
\begin{equation*}
\{\mathbf{L},H_{qm}\}_{{\footnotesize MB}}=0\neq\{\mathbf{L},H\}_{%
{\footnotesize MB}}\;.
\end{equation*}
Also note there will be quantum corrections to the classical equations of
motion for $\left( p_{x},p_{y}\right) $, but not $\left( x,y\right) $.

In phase-space quantization \cite{moyal,cfz,czreview}, the Wigner function
(WF) $f\left( x,p_{x},y,p_{y}\right) $, the Weyl kernel function of the
density operator, evolves according to Moyal's equation \cite{moyal}. 
\begin{equation*}
\frac{\partial f}{\partial t}=\{H_{qm},f\}_{{\footnotesize MB}}\;.
\end{equation*}
In addition to it, the WFs for pure stationary states also satisfy Fairlie's
``$\star$-genvalue'' equations \cite{dbf,cfz} specifying the spectrum. 
\begin{equation*}
H_{qm}(x,p)\star f(x,p)=f(x,p)\star H_{qm}(x,p)=
\end{equation*}%
\begin{equation*}
H_{qm}\left( x+{\frac{i\hbar}{2}}\overset{\rightarrow}{\partial}_{p}~,~p-{%
\frac{i\hbar}{2}}\overset{\rightarrow}{\partial}_{x}\right)
~f(x,p)=E~f(x,p)~.
\end{equation*}
Eigenvalue problems of this type also occur in spectral theory for (special)
Jordan algebras, a point to be re-emphasized below.

The spectrum of this Hamiltonian, then, is proportional to the $\hbar
^{2}l(l+1)$ spectrum of the $so(3)$ Casimir invariant $\mathbf{L}\cdot \star%
\mathbf{L}=L_{+}\star L_{-}+L_{z}\star L_{z}-\hbar L_{z}$, for integer $l$ %
\cite{bffls}. This can be proved algebraically by the identical standard
recursive ladder operations in quantum phase-space which are used in the
operator formalism Fock space, 
\begin{equation*}
L_{z}\star L_{+}-L_{+}\star L_{z}=\hbar L_{+}~,
\end{equation*}
where $L_{\pm}\equiv L_{x}\pm iL_{y}$. From the \emph{real }$\star $\emph{%
-square theorem} \cite{hbar}, it follows that 
\begin{equation*}
\langle\mathbf{L}\cdot\star\mathbf{L}-L_{z}\star L_{z}\rangle=\langle
L_{x}\star L_{x}+L_{y}\star L_{y}\rangle\geq0\;.
\end{equation*}
The $\star$-genvalues of $L_{z}$, $m$, are thus bounded, $|m|\leq l<\sqrt{%
\langle\mathbf{L}\cdot\star\mathbf{L}\rangle}/\hbar$, necessitating $%
L_{-}\star f_{m=-l}=0$. Hence 
\begin{equation*}
L_{+}\star L_{-}\star f_{-l}=0=(\mathbf{L}\cdot\star\mathbf{L}-L_{z}\star
L_{z}+\hbar L_{z})\star f_{-l}~,
\end{equation*}
and consequently $\langle\mathbf{L}\cdot\star\mathbf{L}\rangle=\hbar
^{2}l(l+1)$. Similar $\star$-ladder arguments and inequalities apply
directly in phase space to all Lie algebras.

Classical Hamiltonians are scalars under canonical transformations, but it
should not be assumed that the quantum mechanical expression above is a
canonical scalar \cite{CurtrightZachos}. \ For canonical transformations in
phase-space quantization see \cite{cfz}. \ The $\star$-product and WFs are
also not invariant under canonical transformations, in general, but
transform in a suitable quantum covariant way \cite{cfz}, so as to yield an
identical MB algebra and $\star$-genvalue equations, and thus spectrum,
following from the identical group theoretical construction.

\section{Quantum N-sphere}

For the generic sphere models, $S^{N}$, the maximally symmetric Hamiltonians
are the quadratic Casimir invariants of $so(N+1)$, 
\begin{equation*}
H=\frac{1}{2}P_{a}P_{a}+\frac{1}{4}L_{ab}L_{ab}~,
\end{equation*}
where again in terms of equatorial plane coordinates 
\begin{equation*}
L_{ab}=q^{a}p_{b}-q^{b}p_{a}~,\qquad\qquad P_{a}=\sqrt{1-q^{2}}~p_{a}~,
\end{equation*}
for $a=1,\cdots,N$, $\ q^{2}=\sum_{a=1}^{N}q^{a}q^{a}$ . \ These are\ the
usual angular and de Sitter momenta of $so(N+1)/so(N)$. \ All of these $%
N(N+1)/2$ sphere transformations are symmetries of the classical
Hamiltonian. \ (There are more of them than the $2N-1$ allowed \emph{%
independent} invariants on the phase-space, if $N>2$. \ We will explain how
to select $2N-1$ independent invariants later, when we discuss Nambu
brackets.)

Quantization proceeds as in $S^{2}$, maintaining conservation of all $P_{a}$
and $L_{ab}$, 
\begin{equation*}
H_{qm}=\frac{1}{2}P_{a}\star P_{a}+\frac{1}{4}L_{ab}\star L_{ab}~,
\end{equation*}
and hence the quantum correction is 
\begin{equation*}
H_{qm}-H={\frac{\hbar^{2}}{8}}\left( \frac{1}{1-q^{2}}-1-N(N-1)\right) \;.
\end{equation*}
The spectra are proportional to the Casimir eigenvalues $l(l+N-1)$ for
integer $l$ \cite{bffls}. \ (For $N=3$, this form is reconciled with the
Casimir expression in the next section as $l=2j$, and agrees with \cite%
{lakshmanan,higgs,leemon}).

Can the above quantum Hamiltonian be expressed geometrically through the use
of tangent-space methods? \ The classical description is indeed simple upon
utilization of Vielbeine.%
\begin{equation*}
g_{ab}=\delta_{ij}V_{a}^{i}V_{b}^{j}=\delta_{ab}+\frac{1}{1-q^{2}}%
q^{a}q^{b}\;,\;\;\;g^{ab}=\delta_{ab}-q^{a}q^{b}\;,\;\;%
\;g^{ab}V_{a}^{i}V_{b}^{j}=\delta^{ij}\;.
\end{equation*}
For the generic sphere models, $S^{N}$, standard choices for the Vielbeine
are 
\begin{equation*}
V_{a}^{i}=\delta_{ai}-\frac{q^{a}q^{i}}{q^{2}}\left( 1\pm\frac{1}{\sqrt{%
1-q^{2}}}\right) ,\qquad V^{ai}=\delta_{ai}-\frac{q^{a}q^{i}}{q^{2}}\left(
1\pm\sqrt{1-q^{2}}\right) \;.
\end{equation*}
The classical Hamiltonian equals 
\begin{equation*}
H=\frac{1}{2}(p_{a}V^{ai})(V^{bi}p_{b})\;,
\end{equation*}
but the quantum Hamiltonian is \emph{not} equal to the obvious guess 
\begin{equation*}
H_{other}\equiv\frac{1}{2}(p_{a}V^{ai})\star(V^{bi}p_{b})\neq H_{qm}\;.
\end{equation*}
Although we will see below how this simple form \emph{does} apply in chiral
models, particularly for the $S^{3}$ case, through a different choice of
Dreibeine.

The MBs of cotangent bundle currents, for a general manifold, do not close
among the Vielbein-currents, but instead, 
\begin{equation*}
\left\{ V^{aj}p_{a},V^{bk}p_{b}\right\} _{MB}=\omega^{a[jk]}p_{a}=\left(
V^{bk}\partial_{b}V^{aj}-V^{bj}\partial_{b}V^{ak}\right) p_{a}~,
\end{equation*}
where for the $N$-sphere 
\begin{equation*}
\omega^{a[ij]}=\left( \delta^{ai}q^{j}-\delta^{aj}q^{i}\right)
~w\;,\;\;\;\;\;w\equiv(1-\sqrt{1-q^{2}})/q^{2}\;,
\end{equation*}
choosing the $-$ sign in the definition of the Vielbeine so $%
V^{aj}=\delta_{aj}-q^{a}q^{j}w$. \ It follows that 
\begin{equation*}
H_{qm}-H_{other}=\frac{1}{8}\hbar^{2}\left( N-1\right) \left( 1-2w-N\right)
\;.
\end{equation*}
$H_{other}$ corresponds to a different operator ordering in the conventional
Hilbert space formulation, and has less symmetry than $H_{qm}$. $\ H_{other}$
conserves the rotations $L_{ab}$ (i.e. it is symmetric under the $SO(N)$
stability subgroup for $S^{N}$). \ However, it does not conserve the
Vielbein-currents on the N-sphere, nor does it conserve the de Sitter
momenta. This last statement follows from the above difference $%
H_{qm}-H_{other}$ dictating, 
\begin{equation*}
\{H_{other},P_{c}\}_{{\footnotesize MB}}=\left\{
H_{other}-H_{qm},P_{c}\right\} _{MB}=\hbar^{2}q^{c}(N-1)\left( \frac{2w-1}{%
4q^{2}}\right) \neq0\;.
\end{equation*}

Nevertheless, it can be shown that a $\star$-similarity transformation
compensates for the difference in these Hamiltonians. Consider 
\begin{equation*}
w^{-\frac{(N-1)}{2}}\star p_{a}V^{aj}\star w^{\frac{(N-1)}{2}%
}=w^{1-N}\star\left( p_{a}V^{aj}w^{N-1}\right) =V^{aj}p_{a}-\tfrac{1}{2}%
i\hbar\left( N-1\right) V^{aj}\partial_{a}\ln w,
\end{equation*}
and the complex conjugate transformation 
\begin{equation*}
w^{\frac{(N-1)}{2}}\star p_{a}V^{aj}\star w^{-\frac{(N-1)}{2}}=\left(
w^{N-1}V^{aj}p_{a}\right) \star w^{1-N}=V^{aj}p_{a}+\tfrac{1}{2}i\hbar\left(
N-1\right) V^{aj}\partial_{a}\ln w.
\end{equation*}
Associativity of the $\star$-product then allows the maximally symmetric
real Hamiltonian to be written as 
\begin{equation*}
H_{qm}=\frac{1}{2}\left( w^{N-1}V^{aj}p_{a}\right) \star
w^{-2(N-1)}\star\left( w^{N-1}V^{bj}p_{b}\right) .
\end{equation*}
Using homogeneous coordinates on the sphere, $w=1/\left( 1+\cos\theta\right) 
$, where $\theta$ is the polar angle.

\section{Chiral Models}

The treatment of the 3-sphere $S^{3}$ also accords to the standard chiral
model technology using left- and right-invariant Vielbeine. \ Specifically,
the two choices for such Dreibeine for the 3-sphere are \cite{cuz}: $%
q^{2}=x^{2}+y^{2}+z^{2}$%
\begin{equation*}
^{(\pm)}V_{a}^{i}=\epsilon^{iab}q^{b}\pm\sqrt{1-q^{2}}~g_{ai}~,\qquad
\qquad^{(\pm)}V^{ai}=\epsilon^{iab}q^{b}\pm\sqrt{1-q^{2}}~\delta^{ai}\;.
\end{equation*}
The corresponding right and left conserved charges (left- and
right-invariant, respectively) then are 
\begin{equation*}
\mathcal{R}^{i}=~^{(+)}V_{a}^{i}~\frac{d}{dt}q^{a}=~^{(+)}V^{ai}p_{a}~,%
\qquad\qquad\mathcal{L}^{i}=~^{(-)}V_{a}^{i}~\frac{d}{dt}%
q^{a}=~^{(-)}V^{ai}p_{a}~.
\end{equation*}
More intuitive than those for $S^{2}$ are the linear combinations into Axial
and Isospin charges (again linear in the momenta), 
\begin{equation*}
\tfrac{1}{2}\left( {\mathcal{R}-\mathcal{L}}\right) =\sqrt{1-q^{2}}~\mathbf{p%
}\equiv\mathbf{A\;},\qquad\tfrac{1}{2}\left( {\mathcal{R}+\mathcal{L}}%
\right) =\mathbf{q}\times\mathbf{p}\equiv\mathbf{I\;}.
\end{equation*}
It can easily be seen that the $\mathcal{L}$s and the $\mathcal{R}$s have
PBs closing into standard $su(2)\otimes su(2)$, ie, $su(2)$ relations within
each set, and vanishing between the two sets. Thus they are seen to be
constant, since the Hamiltonian (and also the Lagrangian) can, in fact, be
written in terms of either quadratic Casimir invariant. 
\begin{equation*}
H=\tfrac{1}{2}\mathcal{L}\cdot\mathcal{L}=\tfrac{1}{2}\mathcal{R}\cdot%
\mathcal{R\;}.
\end{equation*}

Quantization consistent with integrability thus proceeds as above for the
2-sphere, since the MB algebra collapses to PBs again, and so the quantum
invariants $\mathcal{L}$ and $\mathcal{R}$ again coincide with the classical
ones, without deformation (quantum corrections). The $\star $-product is now
the obvious generalization to 6-dimensional phase-space. The eigenvalues of
the relevant Casimir invariant are now $j(j+1)$, for half-integer $j$.
However, for this chiral model the symmetric quantum Hamiltonian is simpler,
geometrically, than that of the previous N-sphere, since it can also be
written as 
\begin{equation*}
H_{qm}=\frac{1}{2}(p_{a}V^{ai})\star (V^{bi}p_{b})=\frac{1}{2}\left(
g^{ab}p_{a}p_{b}+{\frac{\hbar ^{2}}{4}}\partial _{a}V^{bi}\partial
_{b}V^{ai}\right) =\tfrac{1}{2}\mathcal{L}\cdot \star \mathcal{L}=\tfrac{1}{2%
}\mathcal{R}\cdot \star \mathcal{R\;}.
\end{equation*}%
No $\star $-similarity transformations are needed for chiral models, unlike
the general N-sphere. \ The Dreibeine throughout this formula can be either $%
~^{+}V_{a}^{i}$ or $~^{-}V_{a}^{i}$, corresponding to either the right- or
the left-acting quadratic Casimir invariant. \ The quantum correction then
amounts to 
\begin{equation*}
H_{qm}-H={\frac{\hbar ^{2}}{8}}(\frac{1}{1-q^{2}}-7)\;.
\end{equation*}%
This expression again is not canonically invariant. Eg, in gnomonic $PR_{N}$
coordinates\footnote{%
The inverse gnomonic Vielbein is polynomial, $\mathsf{V}^{aj}=\delta
^{aj}+Q^{j}Q^{a}+\epsilon ^{jab}Q^{b}$. \ See \cite{higgs}.}, it is $\frac{3%
}{4}\hbar ^{2}(Q^{2}-1)$, ie, it has not transformed as a canonical scalar %
\cite{cfz,CurtrightZachos}. \ Note, however, that either of these quantum
corrections are $\geq -\frac{3}{4}\hbar ^{2}$ on the manifold.

In general, the above discussion also applies to all chiral models, with the
algebra for a chiral group $G$ replacing $su(2)$ above. Ie, the
Vielbein-momenta combinations $V^{aj}p_{a}$ represent algebra generator
invariants, whose quadratic Casimir group invariants yield the respective
Hamiltonians, and whence the properly $\star$-ordered quantum Hamiltonians
as above.

That is to say, for \cite{bcz} group matrices $U$ generated by exponentiated
constant group algebra matrices $T$ weighted by functions of the particle
coordinates $q$, we have 
\begin{equation*}
iU^{-1}\frac{d}{dt}U=~^{(+)}V_{a}^{j}T_{j}\frac{d}{dt}%
q^{a}=~^{(+)}V^{aj}p_{a}T_{j}~,\qquad\qquad iU\frac{d}{dt}%
U^{-1}=~^{(-)}V^{aj}p_{a}T_{j}~,
\end{equation*}
It follows that PBs of left- and right-invariant charges,%
\begin{equation*}
^{(\pm)}V^{aj}p_{a}=\frac{i}{2}Tr\left( T_{j}U^{\mp1}\frac{d}{dt}U^{\pm
1}\right) \;,
\end{equation*}
close to the identical Lie algebras, 
\begin{equation*}
\{^{(\pm)}V^{aj}p_{a},^{(\pm)}V^{bk}p_{b}\}_{{\footnotesize PB}%
}=-2f^{jkn}~^{(\pm)}V^{an}p_{a}~,
\end{equation*}
and PB commute with each other, 
\begin{equation*}
\{~^{(+)}V^{aj}p_{a},^{(-)}V^{bk}p_{b}\}_{{\footnotesize PB}}=0\;.
\end{equation*}
These two statements are implicit in \cite{bcz} as well as throughout the
literature, and are fully explicated in \cite{CurtrightZachos}.

MBs collapse to PBs by linearity in momenta as before, and the Hamiltonian
is again the simple form 
\begin{equation*}
H_{qm}=\frac{1}{2}(p_{a}V^{ai})\star(V^{bi}p_{b})\;.
\end{equation*}
The spectra are given by the Casimir eigenvalues for the relevant algebras
and representations. \ The quantum correction is now found to be 
\begin{equation*}
H_{qm}-H={\frac{\hbar^{2}}{8}}\left(
\Gamma_{ac}^{b}~g^{cd}\Gamma_{bd}^{a}-f_{ijk}f_{ijk}\right) \;,
\end{equation*}
(reducing to the previous result for $S^{3}$). \ In operator language, this
Hamiltonian amounts to an obvious Weyl-ordering of all products \cite%
{CurtrightZachos}.

\section{Nambu Dynamics}

All the models considered above have extra invariants beyond the number of
conserved quantities in involution (mutually commuting) required for
integrability in the Liouville sense. The most systematic way of accounting
for such additional invariants, and placing them all on a more equal
footing, even when they do not all simultaneously commute, is the Nambu
bracket formalism. \ 

\subsection{Classical Nambu Mechanics}

For example, the classical mechanics of a particle on an N-sphere as
discussed above may be summarized elegantly through Nambu mechanics in phase
space \cite{nambu,takhtajan}. Specifically, \cite{hietnambu,nutku}, in an $N$%
-dimensional space, and thus $2N$-dimensional phase space, motion is
confined on the constant surfaces specified by the algebraically independent
integrals of the motion (eg, $L_{x},L_{y},L_{z}$ for $S^{2}$ above.) \
Therefore, the phase-space velocity $\mathbf{v}=(\dot{\mathbf{q}},\dot{%
\mathbf{p}})$ is always perpendicular to the $2N$-dimensional phase-space
gradients $\nabla=(\partial_{\mathbf{q}},\partial_{\mathbf{p}})$ of all
these integrals of the motion.

As a consequence, if there are $2N-1$ algebraically independent such
integrals (ie, the system is maximally superintegrable \cite{winter}), the
phase-space velocity must be proportional \cite{hietnambu} to the
cross-product of all those gradients, and hence the motion is fully
specified for any phase-space function $k(\mathbf{q},\mathbf{p})$ by a
phase-space Jacobian which amounts to the Nambu Bracket. 
\begin{align*}
\frac{dk}{dt} & \equiv\{k,I_{1},\cdots,I_{2N-1}\}_{{\footnotesize NB}}=%
\mathbf{v}\cdot\nabla k \\
& \propto\partial_{i_{1}}k~\epsilon^{i_{1}i_{2}\cdots i_{2N}}~\partial
_{i_{2}}I_{1}\cdots\partial_{i_{2N}}I_{2N-1}=\frac{\partial(k,I_{1},\cdots,%
\cdots,I_{2N-1})}{\partial(q_{1},p_{1},q_{2},p_{2},\cdots,q_{N},p_{N})}\;.
\end{align*}
For instance, consider the above $S^{2}$ case to find the concise result 
\begin{equation*}
\frac{dk}{dt}=\frac{\partial(k,L_{x},L_{y},L_{z})}{\partial(x,p_{x},y,p_{y})}%
~.
\end{equation*}
For the more general $S^{N}$, one now has a choice of $2N-1$ of the $%
N(N+1)/2 $ invariants of $so(N+1)$. \ One of several possible expressions is 
\begin{equation*}
\frac{dk}{dt}=\frac{\left( -1\right) ^{N-1}}{P_{2}P_{3}\cdots P_{N-1}}\frac{%
\partial\left(
k,P_{1},L_{12},P_{2},L_{23},P_{3},\cdots,P_{N-1},L_{N-1\;N},P_{N}\right) }{%
\partial\left( x_{1},p_{1},x_{2},p_{2},\cdots,x_{N},p_{N}\right) }~,
\end{equation*}
where $P_{a}=\sqrt{1-q^{2}}~p_{a}$, for $a=1,\cdots,N$, and $%
L_{a,a+1}=q^{a}p_{a+1}-q^{a+1}p_{a}$, for $~a=1,\cdots,N-1$. 
In general \cite{takhtajan}, classical NBs are Jacobian determinants and
possess all antisymmetries of such. As they are linear in all derivatives,
they also obey the Leibniz rule of partial differentiation. 
\begin{equation*}
\{k(L,M),f_{1},f_{2},\cdots\}_{{\footnotesize NB}}=\frac{\partial k}{%
\partial L}\{L,f_{1},f_{2},\cdots\}_{{\footnotesize NB}}+\frac{\partial k}{%
\partial M}\{M,f_{1},f_{2},\cdots\}_{{\footnotesize NB}}\;.
\end{equation*}
Thus, an entry in the NB algebraically dependent on the rest leads to a
vanishing bracket. For example, it is seen directly from above that the
Hamiltonian is constant, 
\begin{equation*}
\frac{dH}{dt}=\left\{ {\frac{\mathbf{L}\cdot\mathbf{L}}{2}},\cdots\right\}
_{NB}=0\;,
\end{equation*}
since each term of this NB vanishes. This also applies to all explicit
examples discussed here, as they are all maximally superintegrable.

The impossibility to antisymmetrize more than $2N$ indices in $2N$%
-dimensional phase space, 
\begin{equation*}
\epsilon ^{ab\cdots .c[i}\epsilon ^{j_{1}j_{2}\cdots j_{2N}]}=0\;,
\end{equation*}%
leads to the so-called ``fundamental identity'' \cite{takhtajan}, slightly
generalized here. \ For any $f_{i},\;g_{j},\;$and $V,$ 
\begin{align*}
& \{V\{g_{1},\cdots ,g_{2N-1},f_{1}\}_{{\footnotesize NB}},f_{2},\cdots
,f_{2N}\}_{{\footnotesize NB}} \\
& +\{f_{1},V\{g_{1},\cdots ,g_{2N-1},f_{2}\}_{{\footnotesize NB}%
},f_{3},\cdots ,f_{2N}\}_{{\footnotesize NB}} \\
& +\cdots +\{f_{1},\cdots ,f_{2N-1},V\{g_{1},\cdots ,g_{2N-1},f_{2N}\}_{%
{\footnotesize NB}}\}_{{\footnotesize NB}} \\
& =\{g_{1},\cdots ,g_{2N-1},V\{f_{1},f_{2},\cdots ,f_{2N}\}_{{\footnotesize %
NB}}\}_{{\footnotesize NB}}\;.
\end{align*}%
The proportionality factor $V$ in 
\begin{equation*}
\frac{dk}{dt}=V\{k,I_{1},\cdots ,I_{2N-1}\}_{{\footnotesize NB}}\;,
\end{equation*}%
has to be a time-invariant \cite{nutku} if it has no \emph{explicit} time
dependence. \ This is seen from consistency of 
\begin{equation*}
\frac{d}{dt}(V\{f_{1},\cdots ,f_{2N}\}_{{\footnotesize NB}})=\dot{V}%
\{f_{1},\cdots ,f_{2N}\}_{{\footnotesize NB}}+V\{\dot{f_{1}},\cdots
,f_{2N}\}_{{\footnotesize NB}}+\cdots +V\{f_{1},\cdots ,\dot{f}_{2N}\}_{%
{\footnotesize NB}}\;.
\end{equation*}%
where $\dot{V}=\frac{dV}{dt}$, etc. \ This yields 
\begin{multline*}
V\{V\{f_{1},\cdots ,f_{2N}\}_{{\footnotesize NB}},I_{1},\cdots ,I_{2N-1}\}_{%
{\footnotesize NB}}=\dot{V}\{f_{1},\cdots ,f_{2N}\}_{{\footnotesize NB}} \\
+V\{V\{f_{1},I_{1},\cdots ,I_{2N-1}\}_{{\footnotesize NB}},\cdots ,f_{2N}\}_{%
{\footnotesize NB}}+\cdots +V\{f_{1},\cdots ,V\{f_{2N},I_{1},\cdots
,I_{2N-1}\}_{{\footnotesize NB}}\}_{{\footnotesize NB}}\;,
\end{multline*}%
and, by virtue of the FI, $\frac{dV}{dt}=0$ follows \cite{CurtrightZachos}.

Actually, PBs result from a maximal reduction of NBs, by inserting $2N-2$
phase-space coordinates and summing over them, thereby taking \emph{%
symplectic traces}, 
\begin{equation*}
\left\{ L,M\right\} _{PB}=\frac{1}{\left( N-1\right) !}\left\{
L,M,x_{i_{1}},p_{i_{1}},\cdots,x_{i_{N-1}},p_{i_{N-1}}\right\} _{NB}\;,
\end{equation*}
where summation over all $N-1$ pairs of repeated indices is understood.
Fewer traces lead to relations between NBs of maximal rank, $2N$, and those
of lesser rank, $2k$, 
\begin{equation*}
\left\{ L_{1},\cdots,L_{2k}\right\} _{NB}=\frac{1}{\left( N-k\right) !}%
\left\{
L_{1},\cdots,L_{2k},x_{i_{1}},p_{i_{1}},\cdots,x_{i_{N-k}},p_{i_{N-k}}\right%
\} _{NB}\;,
\end{equation*}
(which is one way to define the lower rank NBs for $k\neq1$), or between two
lesser rank NBs. A complete theory of these relations has not been
developed; but, naively, $\left\{ L_{1},\cdots,L_{2k}\right\} _{NB}$ acts
like a Dirac Bracket (DB) up to a normalization, $\{L_{1},L_{2}\}_{DB}$,
where the fixed additional entries $L_{3},\cdots,L_{2k}$ in the NB play the
role of the constraints in the DB. (In effect, this has been previously
observed, e.g., \cite{takhtajan,nutku}, for the extreme case $N=k$, without
symplectic traces.)

By applying this symplectic trace to a general system---not only a
superintegrable one---Hamilton's equations admit an NB expression different
than Nambu's original one, namely 
\begin{equation*}
\frac{dk}{dt}=\{k,H\}_{{\footnotesize PB}}={\frac{1}{(N-1)!}}%
\{k,H,x_{i_{1}},p_{i_{1}},\cdots,x_{i_{N-1}},p_{i_{N-1}}\}_{{\footnotesize NB%
}}\;.
\end{equation*}

\subsection{Quantum Nambu Mechanics}

Despite considerable interest in the problem for nearly 30 years, the
quantization of the Nambu formalism was not completely settled until
recently. \ We believe a transparent, user-friendly technique is now at hand
(see \cite{CurtrightZachos,ZachosCurtright}). \ The problem of quantizing
Nambu brackets \emph{might} be solved by the Abelian deformation method \cite%
{dfst}, but this is not yet clear.

Remarkably, however, the quantization is completely solved by Nambu's
original method, when consistently applied to the phase-space formalism. \
Define quantum Nambu brackets (QNBs) \ 
\begin{equation*}
\left[ A_{1},A_{2},\cdots,A_{k}\right] _{\star}\equiv\sum _{\substack{ \text{%
all }k!\text{\ perms }\left\{ p_{1},p_{2},\cdots ,p_{k}\right\}  \\ \text{of
the indices }\left\{ 1,2,\cdots,k\right\} }}\left( -1\right) ^{\pi\left(
p\right) }A_{p_{1}}\star A_{p_{2}}\star\cdots\star A_{p_{k}}\;.
\end{equation*}
Use these anti-symmetrized $\star$-products in the quantum theory instead of
the previous jacobians. This approach grants in an obvious way only one of
three mathematical desiderata: \ \emph{full antisymmetry}. \ With these
definitions, the Leibniz property and the Fundamental Identity are not
manifestly satisfied. \ But to some extent, the apparent loss of the latter
two properties is a subjective shortcoming, and dependent on the specific
application context.

Even order QNBs may always be resolved into sums of products of commutators: 
$\ \left[ A,B\right] _{\star}=A\star B-B\star A$. \ For instance, 
\begin{align*}
\left[ A,B,C,D\right] _{\star} & =\left[ A,B\right] _{\star}\star\left[ C,D%
\right] _{\star}-\left[ A,C\right] _{\star}\star\left[ B,D\right] _{\star}-%
\left[ A,D\right] _{\star}\star\left[ C,B\right] _{\star} \\
& +\left[ C,D\right] _{\star}\star\left[ A,B\right] _{\star}-\left[ B,D%
\right] _{\star}\star\left[ A,C\right] _{\star}-\left[ C,B\right]
_{\star}\star\left[ A,D\right] _{\star}\;.
\end{align*}
Let us use this fact and re-consider the $S^{2}$ example. \ The result is a
combinatoric identity relating 4 brackets to commutators. 
\begin{equation*}
\left[ A,L_{x},L_{y},L_{z}\right] _{\star}=i\hbar\left[ A,L_{x}\star
L_{x}+L_{y}\star L_{y}+L_{z}\star L_{z}\right] _{\star}\;,
\end{equation*}
as follows from the $su\left( 2\right) $\ MB algebra and the commutator
resolution of the 4-bracket. \ Therefore in this case the Leibniz properties
(and the particular FIs corresponding to them) \emph{are} satisfied, leading
to an effective fundamental identity (EFI). \ Explicitly, 
\begin{equation*}
\left[ A\star B,L_{x},L_{y},L_{z}\right] _{\star}=A\star\left[
B,L_{x},L_{y},L_{z}\right] _{\star}+\left[ A,L_{x},L_{y},L_{z}\right]
_{\star}\star B\;.
\end{equation*}
As a consequence, time evolution of any WF for $S^{2}$\ is given by the QNB
expression: 
\begin{equation*}
\frac{\partial f}{\partial t}=\frac{1}{2\left( i\hbar\right) ^{2}}\left[
L_{x},L_{y},L_{z},f\right] _{\star}\;.
\end{equation*}
In particular, this approach is in agreement with the $\star$-product
quantization of the equations of motion. 
\begin{equation*}
\frac{dx}{dt}=\frac{-1}{2\hbar^{2}}\left[ x,L_{x},L_{y},L_{z}\right]
_{\star}\;,\qquad\qquad\frac{dp_{x}}{dt}=\frac{-1}{2\hbar^{2}}\left[
p_{x},L_{x},L_{y},L_{z}\right] _{\star}\;,
\end{equation*}
where the second of these does indeed incorporate the previously given
quantum correction.

How does the use of 4-brackets extend to other examples? \ Any Lie algebra
will allow a commutator with a quadratic Casimir to be rewritten as a \emph{%
sum} of 4-brackets. \ Suppose 
\begin{equation*}
\left[ Q_{a},Q_{b}\right] _{\star }=i\hbar f_{abc}Q_{c}\;,
\end{equation*}%
in a basis where $f_{abc}$ is totally antisymmetric (this particular choice
of basis is helpful, but not crucial). \ Then for a \emph{sum} (over all
repeated indices) of quantum Nambu 4-brackets we have 
\begin{equation*}
f_{abc}\left[ A,Q_{a},Q_{b},Q_{c}\right] _{\star }=\left[ A,f_{abc}\left[
Q_{a},Q_{b},Q_{c}\right] _{\star }\right] _{\star }\;.
\end{equation*}%
Only a commutator with the trilinear invariant survives. \ Moreover, this
trilinear invariant reduces to the quadratic Casimir. 
\begin{equation*}
f_{abc}\left[ Q_{a},Q_{b},Q_{c}\right] _{\star }=3f_{abc}Q_{a}\star \left[
Q_{b},Q_{c}\right] _{\star }=3i\hbar f_{abc}f_{bcd}Q_{a}\star Q_{d}\;.
\end{equation*}%
For simple Lie algebras (with appropriately normalized charges) we have 
\begin{equation*}
f_{abc}f_{bcd}=c_{\text{adjoint}}\,\delta _{ad}\;,
\end{equation*}%
where $c_{\text{adjoint}}$ is a \emph{number}. \ For example, $c_{\text{%
adjoint}}=N$ for $su\left( N\right) $. \ 

So we obtain the Casimir $Q_{a}\star Q_{a}$%
\begin{equation*}
f_{abc}\left[ Q_{a},Q_{b},Q_{c}\right] _{\star}=3i\hbar\,c_{\text{adjoint}%
}\,Q_{a}\star Q_{a}\;,
\end{equation*}
and we conclude that 
\begin{equation*}
f_{abc}\left[ A,Q_{a},Q_{b},Q_{c}\right] _{\star}=3i\hbar\,c_{\text{adjoint}%
}\,\left[ A,Q_{a}\star Q_{a}\right] _{\star}\;.
\end{equation*}
As a corollary, we again have the effective 4-bracket fundamental identity
(EFI) 
\begin{multline*}
f_{abc}\left[ Q_{a},Q_{b},Q_{c},\left[ A,B,\cdots,D\right] _{\star}\right]
_{\star} \\
=f_{abc}\left[ \left[ Q_{a},Q_{b},Q_{c},A\right] _{\star},B,\cdots ,D\right]
_{\star}+f_{abc}\left[ A,\left[ Q_{a},Q_{b},Q_{c},B\right] _{\star},\cdots,D%
\right] _{\star}+\cdots \\
+f_{abc}\left[ A,B,\cdots,\left[ Q_{a},Q_{b},Q_{c},D\right] _{\star }\right]
_{\star}\;.
\end{multline*}
By using this result, \emph{all} the models above can be quantized through
the use of QNBs to describe the time evolution of their WFs.\ 

More than three invariants may also be incorporated into a QNB. \ Only a
careful and complete physical interpretation of the result is needed in the
general case. \ For example, for $S^{N}$ with $N>2$ a full set of $2N-1$
invariants leads from the previously given classical jacobian to a QNB: 
\begin{align*}
\frac{\partial\left(
f,P_{1},L_{12},P_{2},\cdots,P_{N-1},L_{N-1\;N},P_{N}\right) }{\partial\left(
x_{1},p_{1},x_{2},p_{2},\cdots,x_{N},p_{N}\right) } & =\left( -1\right)
^{N-1}P_{2}\cdots P_{N-1}\frac{df}{dt} \\
& =\left( -1\right) ^{N-1}\frac{d}{dt}\left( P_{2}\cdots P_{N-1}f\right) \;,
\end{align*}
becomes, for $\hbar\neq0$, 
\begin{align*}
\frac{1}{\left( i\hbar\right) ^{N}N}\left[ f,P_{1},L_{12},P_{2},%
\cdots,P_{N-1},L_{N-1\;N},P_{N}\right] _{\star} & =\left( -1\right) ^{N-1}%
\frac{d}{dt}\left\{ P_{2},\cdots,P_{N-1},f\right\} _{\star} \\
& +\;\;\text{\emph{quantum-connection terms\ .}}
\end{align*}
The invariants appearing in the totally symmetric generalization of the
Jordan product \cite{Jordan,Kurosh},%
\begin{equation*}
\left\{ A_{1},A_{2},\cdots,A_{k}\right\} _{\star}\equiv\sum _{\substack{ 
\text{all }k!\text{\ perms }\left\{ p_{1},p_{2},\cdots ,p_{k}\right\}  \\ 
\text{of the indices }\left\{ 1,2,\cdots,k\right\} }}A_{p_{1}}\star
A_{p_{2}}\star\cdots\star A_{p_{k}}\;,
\end{equation*}
on the above RHS, 
\begin{equation*}
\left\{ P_{2},P_{3},\cdots,P_{N-1},\frac{df}{dt}\right\} _{\star}=\frac{d}{dt%
}\left\{ P_{2},P_{3},\cdots,P_{N-1},f\right\} _{\star}\;,
\end{equation*}
will have the effect to set, through their net (but \emph{not} simultaneous,
individual) eigenvalues, a variable, dynamical time scale for evolution of
the various eigen-WFs. \ 

In addition, the ``quantum-connection terms'' represent higher order
corrections, in powers of $\hbar$, corresponding to group rotations of $f$
that will be described fully elsewhere. \ For example, if $f$ is the
bilinear 
\begin{equation*}
f_{ab}\equiv\left( L_{a}+P_{a}\right) \star\left( L_{b}-P_{b}\right) \;.
\end{equation*}
then $df_{ab}/dt=0$ for a particle moving freely on the surface of the
3-sphere, but the corresponding group rotation is \emph{not} zero. \ Hence
the six bracket reduces entirely to these quantum connection terms. \
Explicitly we find 
\begin{equation*}
\left[ f_{ab},P_{x},L_{z},P_{y},L_{x},P_{z}\right] _{\star}=4i\hbar^{5}%
\sum_{c}\left( \varepsilon_{b2c}f_{ac}-\varepsilon_{a2c}f_{cb}\right) \;.
\end{equation*}
Note that the rotation on the RHS here is a quantum effect, and vanishes in
the classical limit as given by 
\begin{equation*}
\lim_{\hbar\rightarrow0}\left[ f_{ab},P_{x},L_{z},P_{y},L_{x},P_{z}\right]
_{\star}/\hbar^{3}=0\;.
\end{equation*}

We believe such a complete physical interpretation explains the perceived
failure of the Leibniz rules and FI in the general situation, in a
transparent way, and is the only additional ingredient required for a
successful \emph{non-Abelian} quantum implementation of the most general
Nambu brackets. \ \textit{A priori}, this approach could conceivably be
equivalent to the Abelian deformation approach \cite{dfst}, but no one has
shown this.

These points are worth re-emphasizing for the 3-sphere expressed in the
chiral language. \ Again, let us use $\mathcal{L}_{i}$ and $\mathcal{R}_{i}$ 
$\left( i=1,2,3\sim x,y,z\right) $ for the mutually commuting $su\left(
2\right) $ charges. \ 
\begin{equation*}
\left[ \mathcal{L}_{i},\mathcal{L}_{j}\right] _{\star}=i\hbar\varepsilon
_{ijk}\mathcal{L}_{k}\;,\;\;\;\left[ \mathcal{R}_{i},\mathcal{R}_{j}\right]
_{\star}=i\hbar\varepsilon_{ijk}\mathcal{R}_{k}\;,\;\;\;\left[ \mathcal{L}%
_{i},\mathcal{R}_{j}\right] _{\star}=0\;.
\end{equation*}
Define the usual quadratic Casimirs for the left and right $su\left(
2\right) $'s: 
\begin{equation*}
I_{\mathcal{L}}=\sum_{i}\mathcal{L}_{i}\star\mathcal{L}_{i}\;,\;\;\;I_{%
\mathcal{R}}=\sum_{i}\mathcal{R}_{i}\star\mathcal{R}_{i}\;.
\end{equation*}
We find the simple result: 
\begin{align*}
\left[ f,F\left( I_{\mathcal{L}},I_{\mathcal{R}}\right) ,\mathcal{R}_{x},%
\mathcal{R}_{y},\mathcal{L}_{x},\mathcal{L}_{y}\right] _{\star} & =\left(
i\hbar\right) ^{2}\left\{ \left[ f,F\left( I_{\mathcal{L}},I_{\mathcal{R}%
}\right) \right] _{\star},\mathcal{L}_{z},\mathcal{R}_{z}\right\} _{\star} \\
& =\left( i\hbar\right) ^{2}\left[ \left\{ f,\mathcal{L}_{z},\mathcal{R}%
_{z}\right\} _{\star},F\left( I_{\mathcal{L}},I_{\mathcal{R}}\right) \right]
_{\star}\;,
\end{align*}
where $F\left( I_{\mathcal{L}},I_{\mathcal{R}}\right) $ is \emph{any} $\star$%
\emph{-function} of the left and right Casimirs. \ Physically, this Nambu
bracket is simply time evolution generated by the Hamiltonian $H\equiv
F\left( I_{\mathcal{L}},I_{\mathcal{R}}\right) $ with the Jordan-like
eigenvalues $\sigma,$ of 
\begin{equation*}
\left\{ f,\mathcal{L}_{z},\mathcal{R}_{z}\right\} _{\star}=\sigma f\;,
\end{equation*}
setting the time scales for the various sectors of the theory: \ i.e. \emph{%
dynamical time scales}. \ In particular this is true for the choices $%
F\left( I_{\mathcal{L}},I_{\mathcal{R}}\right) =I_{\mathcal{L}}$ or $F\left(
I_{\mathcal{L}},I_{\mathcal{R}}\right) =I_{\mathcal{R}}$, choices relevant
for the particle on the 3-sphere.

A complete set of solutions consists of all WFs of the form $f_{\lambda
_{1}\rho_{1},\lambda_{2}\rho_{2}}$ where 
\begin{align*}
\mathcal{L}_{z}\star f_{\lambda_{1}\rho_{1},\lambda_{2}\rho_{2}} &
=\lambda_{1}f_{\lambda_{1}\rho_{1},\lambda_{2}\rho_{2}}\;,\;\;\;f_{\lambda
_{1}\rho_{1},\lambda_{2}\rho_{2}}\star\mathcal{L}_{z}=\lambda_{2}f_{%
\lambda_{1}\rho_{1},\lambda_{2}\rho_{2}}\;, \\
\mathcal{R}_{z}\star f_{\lambda_{1}\rho_{1},\lambda_{2}\rho_{2}} & =\rho
_{1}f_{\lambda_{1}\rho_{1},\lambda_{2}\rho_{2}}\;,\;\;\;f_{\lambda_{1}\rho
_{1},\lambda_{2}\rho_{2}}\star\mathcal{R}_{z}=\rho_{2}f_{\lambda_{1}\rho
_{1},\lambda_{2}\rho_{2}}\;.
\end{align*}
Hence $\left\{ f,\mathcal{L}_{z},\mathcal{R}_{z}\right\}
_{\star}=\sigma_{12}f$ with 
\begin{equation*}
\sigma_{12}=2\lambda_{1}\rho_{1}+\lambda_{1}\rho_{2}+\rho_{1}\lambda
_{2}+2\lambda_{2}\rho_{2}\;,
\end{equation*}
since 
\begin{equation*}
\left\{ f,\mathcal{L}_{z},\mathcal{R}_{z}\right\} _{\star}=\left\{ \mathcal{L%
}_{z},\mathcal{R}_{z}\right\} _{\star}\star f+\mathcal{L}_{z}\star f\star%
\mathcal{R}_{z}+\mathcal{R}_{z}\star f\star\mathcal{L}_{z}+f\star\left\{ 
\mathcal{L}_{z},\mathcal{R}_{z}\right\} _{\star}\;.
\end{equation*}
So the time scale is indeed dynamical, and given by $\sigma_{12}$. \ 

The simple Leibniz rule for generic $f$ and $g,$ that would equate 
\begin{align*}
& \left[ f\star g,F\left( I_{\mathcal{L}},I_{\mathcal{R}}\right) ,\mathcal{R}%
_{x},\mathcal{R}_{y},\mathcal{L}_{x},\mathcal{L}_{y}\right] _{\star}\text{ \
\ \ \ and} \\
& f\star\left[ g,F\left( I_{\mathcal{L}},I_{\mathcal{R}}\right) ,\mathcal{R}%
_{x},\mathcal{R}_{y},\mathcal{L}_{x},\mathcal{L}_{y}\right] _{\star}+\left[
f,F\left( I_{\mathcal{L}},I_{\mathcal{R}}\right) ,\mathcal{R}_{x},\mathcal{R}%
_{y},\mathcal{L}_{x},\mathcal{L}_{y}\right] _{\star}\star g\;,
\end{align*}
will fail for products $f_{\lambda_{1}\rho_{1},\lambda_{2}\rho_{2}}\star
g_{\lambda_{2}\rho_{2},\lambda_{3}\rho_{3}}$ unless 
\begin{equation*}
\sigma_{12}=\sigma_{23}=\sigma_{13}\;.
\end{equation*}

There is no quantum-connection term in this particular case due to our
choice for the invariants in the bracket $\left[ f,F\left( I_{\mathcal{L}%
},I_{\mathcal{R}}\right) ,\mathcal{R}_{x},\mathcal{R}_{y},\mathcal{L}_{x},%
\mathcal{L}_{y}\right] _{\star}$. \ The more general situation is revealed
by a different choice as follows: 
\begin{equation*}
\left[ f,\mathcal{L}_{x},\mathcal{L}_{y},\mathcal{L}_{z},\mathcal{R}_{x},%
\mathcal{R}_{y}\right] _{\star}=\tfrac{3}{2}\left( i\hbar\right) ^{2}\left[
\left\{ f,\mathcal{R}_{z}\right\} _{\star},I_{\mathcal{L}}\right] _{\star}+%
\tfrac{1}{2}\left( i\hbar\right) ^{2}\sum_{i}\left[ \,\left[ \left[ f,%
\mathcal{L}_{i}\right] _{\star},\mathcal{L}_{i}\right] _{\star}\,,\,\mathcal{%
R}_{z}\,\right] _{\star}\;,
\end{equation*}
or equivalently 
\begin{equation*}
\left[ f,\mathcal{R}_{x},\mathcal{R}_{y},\mathcal{R}_{z},\mathcal{L}_{x},%
\mathcal{L}_{y}\right] _{\star}=\tfrac{3}{2}\left( i\hbar\right) ^{2}\left[
\left\{ f,\mathcal{L}_{z}\right\} _{\star},I_{\mathcal{R}}\right] _{\star}+%
\tfrac{1}{2}\left( i\hbar\right) ^{2}\sum_{i}\left[ \,\left[ \left[ f,%
\mathcal{R}_{i}\right] _{\star},\mathcal{R}_{i}\right] _{\star}\,,\,\mathcal{%
L}_{z}\,\right] _{\star}\;.
\end{equation*}
These are the exact results for these QNBs. \ The first terms (single
commutators) on the RHS's are inherently $O\left( \hbar^{3}\right) $, and
encode the expected time evolution for Nambu quantum dynamics, while the
second terms (triple commutators) are $O\left( \hbar^{5}\right) $. \ One is
tempted to interpret these extra terms as some type of covariant completion
for this particular example of Nambu time evolution, with ``quantum
connections'' as the given higher order effects in $\hbar$. \ 

As a test case, again take $f$ to be a bilinear $f_{ab}\equiv\mathcal{L}%
_{a}\star\mathcal{R}_{b}$ of specific left and right charges. \ Since $\star$
commutators are indeed $\star$\ derivations, all $\star$\ functions of the
six possible $\mathcal{L}_{a}$ and $\mathcal{R}_{b}$ charges commute with
the Casimirs, so the first terms on the RHS's of the last two equations
vanish for $f=f_{ab}$ (i.e. $f_{ab}$ for a particle moving freely on the
surface of a 3-sphere has no time derivatives). \ However, the second terms
on the RHS's do \emph{not} vanish for $f=f_{ab}$ but are just rotations of
the $\mathcal{L}_{a}$ and $\mathcal{R}_{b}$ charges about the $z$ axis. 
\begin{align*}
\sum_{i}\left[ \,\left[ \left[ f_{ab},\mathcal{L}_{i}\right] _{\star },%
\mathcal{L}_{i}\right] _{\star}\,,\,\mathcal{R}_{z}\,\right] _{\star} &
=2i\hbar^{3}\sum_{c}\varepsilon_{b3c}f_{ac}\;. \\
\sum_{i}\left[ \,\left[ \left[ f_{ab},\mathcal{R}_{i}\right] _{\star },%
\mathcal{R}_{i}\right] _{\star}\,,\,\mathcal{L}_{z}\,\right] _{\star} &
=2i\hbar^{3}\sum_{c=1,2,3}\varepsilon_{a3c}f_{cb}\;.
\end{align*}
For this particular example, then, the total effects of the chosen
invariants in the 6-brackets are 
\begin{align*}
\left[ f_{ab},\mathcal{L}_{x},\mathcal{L}_{y},\mathcal{L}_{z},\mathcal{R}%
_{x},\mathcal{R}_{y}\right] _{\star} & =-i\hbar^{5}\sum_{c}\varepsilon
_{b3c}f_{ac}\;, \\
\left[ f_{ab},\mathcal{R}_{x},\mathcal{R}_{y},\mathcal{R}_{z},\mathcal{L}%
_{x},\mathcal{L}_{y}\right] _{\star} & =-i\hbar^{5}\sum_{c}\varepsilon
_{a3c}f_{cb}\;.
\end{align*}

It remains to link up the QNB approach with the $\star$ eigenvalue equation
for static WFs, as needed to develop the spectral theory in such a
formalism. \ This must be described in detail elsewhere. \ However an
essential point is contained in the previous formulas. \ A complete spectral
theory in the NB framework requires solving totally symmetrized bracket
eigenvalue equations, of the form 
\begin{equation*}
\lambda f=\left\{ I_{1},I_{2},\cdots,I_{k},f\right\} _{\star}\;,
\end{equation*}
for the allowed \emph{net} eigenvalues $\lambda$, where in general $\left[
I_{i},I_{j}\right] _{\star}\neq0$. \ These net eigenvalue equations set the
different time scales under NB evolution as expressed above. \ They are
natural generalizations of the anticommutator eigenvalue equations
encountered in the standard spectral theory of (special) Jordan algebras %
\cite{MoreJordan,JordanVonNeumannWigner}. \ Nevertheless, we have not found
a complete discussion of such equations in the literature.

\paragraph{\textbf{Acknowledgments}}

We thank Y Nambu and Y Nutku for helpful discussions. \ We also thank the
organizers of the superintegrability workshop, especially P Winternitz, for
the opportunity to present this research. \ The warm hospitality of the
Centre de recherches math\'{e}matiques, Universit\'{e} de Montr\'{e}al, is
most gratefully acknowledged. \ 

\section*{Appendix: \ Nambu quantum mechanics of the isotropic oscillator}

The analysis above can also be carried out using conventional Hilbert space
operator techniques. \ Rather than repeat our previous discussion, we take
another example to illustrate operator methods. (For a thorough discussion
of the ideas in this Appendix, see \cite{CQNM}. \ This later paper contains
a more complete list of references.) \ 

Consider the $n$ dimensional oscillator using the standard lowering/raising
operator basis that gives the commutator algebra 
\begin{equation*}
\left[ a_{i},b_{j}\right] =\hbar\delta_{ij}\;,\;\;\;\left[ a_{i},a_{j}\right]
=0=\left[ b_{i},b_{j}\right] \;.
\end{equation*}
Construct the usual bilinear charges that realize the $u\left( n\right) $
algebra 
\begin{equation*}
N_{ij}\equiv b_{i}a_{j}\;,\;\;\;\left[ N_{ij},N_{kl}\right] =\hbar\left(
N_{il}\delta_{jk}-N_{kj}\delta_{il}\right) \;.
\end{equation*}
Then the isotropic Hamiltonian is 
\begin{equation*}
H=\omega\sum_{i=1}^{n}\left( N_{i}+\frac{1}{2}\hbar\right)
\;,\;\;\;N_{i}\equiv N_{ii}\;,
\end{equation*}
which gives the $n^{2}$ conservation laws $\left[ H,N_{ij}\right] =0$. \ Now
when we consider the isotropic oscillator dynamics using quantum Nambu
brackets (QNBs) we obtain the main result for oscillator $2n$-brackets in
the form of a theorem.

\paragraph{\textbf{Isotropic Oscillator Theorem (reductio ad dimidium):}}

Let $\mathsf{N}=N_{1}+N_{2}+\cdots+N_{n}$ and intercalate the $n-1$
non-diagonal operators $N_{i\;i+1}$ for $i=1,\cdots,n-1$ into a $2n$-bracket
with the $n$ mutually commuting $N_{j}$ for $j=1,\cdots,n$ to find 
\begin{align*}
\left[ f,N_{1},N_{12},N_{2},N_{23},\cdots,N_{n-1},N_{n-1\;n},N_{n}\right] &
=\hbar^{n-1}\left\{ \left[ f,\mathsf{N}\right] ,N_{12},N_{23},\cdots,N_{n-1%
\;n}\right\} \\
& =\hbar^{n-1}\left[ \left\{ f,N_{12},N_{23},\cdots,N_{n-1\;n}\right\} ,%
\mathsf{N}\right] \;.
\end{align*}
Here we have used a fully symmetrized Jordan operator product \cite{Jordan}
as generalized by Kurosh \cite{Kurosh} 
\begin{equation*}
\left\{ A_{1},A_{2},\cdots,A_{k}\right\} \equiv\sum_{\substack{ \text{all }k!%
\text{\ perms }\left\{ p_{1},p_{2},\cdots,p_{k}\right\}  \\ \text{of the
indices }\left\{ 1,2,\cdots,k\right\} }}A_{p_{1}}A_{p_{2}}\cdots A_{p_{k}}\;.
\end{equation*}
This was introduced by Pascual Jordan, in the bilinear form, to render
non-Abelian algebras into Abelian algebras at the expense of
non-associativity. \ We have also used the fully antisymmetrized Nambu
operator product, or QNB \cite{nambu}, 
\begin{equation*}
\left[ A_{1},A_{2},\cdots,A_{k}\right] \equiv\sum_{\substack{ \text{all }k!%
\text{\ perms }\left\{ p_{1},p_{2},\cdots,p_{k}\right\}  \\ \text{of the
indices }\left\{ 1,2,\cdots,k\right\} }}\left( -1\right) ^{\pi\left(
p\right) }A_{p_{1}}A_{p_{2}}\cdots A_{p_{k}}\;.
\end{equation*}
The non-diagonal operators $N_{i\;i+1}$ do not all commute among themselves,
nor with all the $N_{j}$, but their non-Abelian properties are encountered
in the above Jordan and Nambu products in a rather minimal way. \ Only
adjacent entries in the list $N_{12},N_{23},N_{34},\cdots,N_{n-1\;n}$ fail
to commute. \ Also in the list of $2n-1$ generators within the original QNB,
each $N_{j}$ fails to commute only with the adjacent $N_{j-1\;j}$ and $%
N_{j\;j+1}$. \ Such a list of invariants constitutes a ``Hamiltonian path''
through the algebra\footnote{%
There are other Hamiltonian paths through the algebra. \ A different set of $%
2n-1$ invariants which leads to an equivalent theorem can be obtained just
by taking an arbitrarily ordered list of the mutually commuting $N_{i}$, and
then intercalating non-diagonal generators to match adjacent indices on the $%
N_{i}$. \ That is, for any permutation of the indices $\left\{
p_{1},\cdots,p_{n}\right\} $, we have: \ $\left[
f,N_{p_{1}},N_{p_{1}p_{2}},N_{p_{2}},N_{p_{2}p_{3}},N_{p_{3}},%
\cdots,N_{p_{n-1}},N_{p_{n-1}\;p_{n}},N_{p_{n}}\right] =\hbar^{n-1}\left\{ %
\left[ f,\mathsf{N}\right] ,N_{p_{1}p_{2}},N_{p_{2}p_{3}},\cdots,N_{p_{n-1}%
\;p_{n}}\right\} =\hbar^{n-1}\left[ \left\{
f,N_{p_{1}p_{2}},N_{p_{2}p_{3}},\cdots,N_{p_{n-1}\;p_{n}}\right\} ,\mathsf{N}%
\right] \;$.}.

\paragraph{\textbf{Proof of the Oscillator Theorem}: \ }

Linearity in each argument and total antisymmetry of the Nambu bracket allow
us to replace any one of the $N_{i}$ by the sum $\mathsf{N}$. \ We choose to
replace $N_{n}\rightarrow\mathsf{N}$, hence to obtain 
\begin{equation*}
\left[ f,N_{1},N_{12},N_{2},\cdots,N_{n-1},N_{n-1\;n},N_{n}\right] =\left[
f,N_{1},N_{12},N_{2},\cdots,N_{n-1},N_{n-1\;n},\mathsf{N}\right] \;.
\end{equation*}
Now since $\left[ \mathsf{N},N_{ij}\right] =0$, the commutator resolution of
the $2n$-bracket implies that $\mathsf{N}$ must appear ``locked'' in a
commutator with $f$, and therefore $f$ cannot appear in any other
commutator. \ But then $N_{1}$ commutes with all the remaining free $N_{ij}$
except $N_{12}$. \ So $N_{1}$ must be locked in $\left[ N_{1},N_{12}\right] $%
. \ Continuing in this way, $N_{2}$ must be locked in $\left[ N_{2},N_{23}%
\right] $, etc., until finally $N_{n-1}$ is locked in $\left[
N_{n-1},N_{n-1\;n}\right] $. \ Thus all $2n$ entries have been paired and
locked in the indicated $n$ commutators, i.e. they are all ``zipped-up''. \
Moreover, these $n$ commutators can and will appear as products ordered in
all $n!$ possible ways with coefficients $+1$ since interchanging a pair of
commutators requires interchanging two pairs of the original entries in the
bracket. \ Thus we conclude 
\begin{equation*}
\left[ f,N_{1},N_{12},N_{2},\cdots,N_{n-1},N_{n-1\;n},N_{n}\right] =\left\{ %
\left[ f,\mathsf{N}\right] ,\left[ N_{1},N_{12},\right] ,\cdots,\left[
N_{n-1},N_{n-1\;n}\right] \right\} \;.
\end{equation*}
Now all the paired $N_{ij}$ commutators evaluate as $\left[
N_{i-1},N_{i-1\;i}\right] =\hbar N_{i-1\;i}$, so we have 
\begin{equation*}
\left[ f,N_{1},N_{12},N_{2},\cdots,N_{n-1},N_{n-1\;n},N_{n}\right]
=\hbar^{n-1}\left\{ \left[ f,\mathsf{N}\right] ,N_{12},\cdots
,N_{n-1\;n}\right\} \;.
\end{equation*}
Finally the commutator with $\mathsf{N}$ may be performed either before or
after the Jordan product of $f$ with all the $N_{i-1\;i}$, since again $%
\left[ \mathsf{N},N_{ij}\right] =0$. \ Hence 
\begin{equation*}
\left\{ \left[ f,\mathsf{N}\right] ,N_{12},\cdots,N_{n-1\;n}\right\} =\left[
\left\{ f,N_{12},\cdots,N_{n-1\;n}\right\} ,\mathsf{N}\right]
\;.\;\;\;\;\blacksquare
\end{equation*}

The oscillator theorem\ is a remarkable relation. \ The invariants which are
in involution (i.e. the Cartan subalgebra of $u\left( n\right) $) are
separated out of the QNB into a single commutator involving their sum, hence 
$H/\omega $, to yield a time derivative, while the invariants which do not
commute ($n-1$ of them, corresponding in number to the rank of $su\left(
n\right) $) are swept into a Jordan product. \ As in the N-sphere examples,
this helps to clarify why the Leibniz rules fail when time evolution is
expressed using QNBs. \ As is fairly well-known, evolution under the QNB
does not satisfy the trivial Leibniz property: $\left[
fg,N_{1},N_{12},N_{2},\cdots ,N_{n-1\;n},N_{n}\right] $ $\neq $ $f\left[
g,N_{1},N_{12},N_{2},\cdots ,N_{n-1\;n},N_{n}\right] $ $+$ $\left[
f,N_{1},N_{12},N_{2},\cdots ,N_{n-1\;n},N_{n}\right] g$. \ But here the
failure of this Leibniz rule for the Nambu bracket has been linked to the
intervention of a generalized Jordan product involving non-commuting
invariants.

\end{document}